\documentclass[%
 reprint,superscriptaddress,nofootinbib, amsmath,amssymb, aps, nofootinbib]{revtex4-1}
\usepackage[sort&compress]{natbib}
\usepackage{graphicx}
\usepackage{dcolumn}
\usepackage{bm}
\usepackage{float}
\usepackage{ragged2e} 
\usepackage{changes}
\usepackage{color}
\usepackage{varioref}
\usepackage{hyperref}
\usepackage{cleveref}
\usepackage{amsmath}
\usepackage{stackengine}

\hypersetup{colorlinks,citecolor=black,filecolor=black,linkcolor=black,urlcolor=black}
\usepackage{stackengine}
\stackMath
\usepackage{textcomp}
\begin{document}
\title{Deep subwavelength thermal switch via resonant coupling in monolayer hexagonal boron nitride}

\author{Georgia T. Papadakis}
\affiliation{Department of Electrical Engineering, Ginzton Laboratory, Stanford University, California, 94305, USA}
\author{Christopher J. Ciccarino}\thanks{These two authors contributed equally}
\affiliation{Harvard John A. Paulson School of Engineering and Applied Sciences, Harvard University, MA}
\author{Lingling Fan}\thanks{These two authors contributed equally}
\affiliation{Department of Electrical Engineering, Ginzton Laboratory, Stanford University, California, 94305, USA}
\author{Meir Orenstein}%
\affiliation{Department of Electrical Engineering, Technion-Israel Institute of Technology, 32000 Haifa, Israel}
\author{Prineha Narang}
\affiliation{Harvard John A. Paulson School of Engineering and Applied Sciences, Harvard University, MA}
\author{Shanhui Fan}\email{shanhui@stanford.edu}
\affiliation{Department of Electrical Engineering, Ginzton Laboratory, Stanford University, California, 94305, USA}

\date{\today}
\begin{abstract}
 \begin{center}
 \end{center}
{Unlike the electrical conductance that can be widely modulated within the same material even in deep nanoscale devices, tuning the thermal conductance within a single material system or nanostructure is extremely challenging and requires a large-scale device. This prohibits the realization of robust ON/OFF states in switching the flow of thermal currents. Here, we present the theory of a thermal switch based on resonant coupling of three photonic resonators, in analogy to the field-effect electronic transistor composed of a source, gate, and drain. As a material platform, we capitalize on the extreme tunability and low-loss resonances observed in the dielectric function of monolayer hexagonal boron nitride (hBN) under controlled strain. We derive the dielectric function of hBN from first principles, including the phonon-polariton linewidths computed by considering phonon-isotope and anharmonic phonon-phonon scattering. Subsequently, we propose a strain-controlled hBN-based thermal switch that modulates thermal conductance by more than an order of magnitude, corresponding to an ON/OFF contrast ratio of $98\%$, in a deep subwavelength nanostructure.}

\begin{description}\item[Keywords] hexagonal boron nitride, coupled-mode theory, fluctuational electrodynamics, near-field\end{description}

\end{abstract}

\maketitle

\section{\label{Intro}INTRODUCTION}

\par{Controlling the flow of a thermal current is of critical importance in all applications that require thermal regulation and efficient dissipation of heat~\cite{HeatDissipation_Review,Heat_Dissipation2009}, as well as in energy harvesting systems~\cite{Lipson_TPV, Yablonovich_PNASTPV2019, Papadakis_TPV2020,Zhang_NearfieldTPV}, and thermal circuitry~\cite{Li_ThermalTransistor, Abdallah_ThermalTransistor}. To achieve this, one ought to gain active control over the thermal conductance of materials. Previous efforts to modulate thermal conductance have focused largely on tailoring the propagation characteristics of acoustic phonons. With this approach, the long wavelength of acoustic phonons restricts the down scaling of thermal modulators to the order of micrometers \cite{Phononics_LengthScaleReview,Phononics_LengthScale1,Phononics_LengthScale2}. Various physical effects have been exploited including temperature-dependent material properties~\cite{VO2_thermal_conductivity}, defects and guest ions~\cite{Graphene_thermal_conductivity_Defects,MoS2_thermal_conductivity_Defects,MoS2_thermal_conductivity}, and mechanical deformation~\cite{MoS2_thermal_conductivity_Defects,polymers_thermal_conductivity_Defects} in materials like polymers, thermoelectrics~\cite{Thermoelectrics_thermal_conductivity}, and phase-change materials~\cite{Phase_transition_thermal_conductivity,VO2_thermal_conductivity}, as well as in composite nanostructures~\cite{Choe_VO2_thermal_conductivity2017}. Despite previous efforts, however, reported contrast ratios of thermal conductance modulation remain well below an order of magnitude, prohibiting the clear distinction between ON and OFF thermal states.}

\par{In contrast, the electrical conductivity can be tuned by more than ten orders of magnitude within the same material system even for deep nano-devices ~\cite{Javey_MoS2_Transistor,ashcroft1976solid}. This allows excellent control of electrical currents and has led to the ubiquitous electronic transistor~\cite{Bardeen_Transistor}. A field-effect transistor (FET) serves as a switch of electric currents based on electrostatic modulation of a nanoscale conductive channel (gate), a functionality that lies at the cornerstone of modern optoelectronics. It is highly desirable to gain better control of the thermal conductance in order to achieve similar degree of control in the flow of heat in the nanoscale.}

\par{Here, we propose a nanoscale thermal switch that can modulate thermal conductance by more than an order of magnitude, thereby achieving an ON/OFF contrast ratio of $98\%$. Our switch consists of three resonators that serve similar functionalities as the source, gate, and drain of its electronic counterpart, the FET. We show that thermal modulation can occur in a deep subwavelength configuration over a transfer distance of few tens of nanometers. To achieve this, the source and drain do not have material contact, thus the transfer mechanism is radiative (photons, or electrons in vacuum). Whereas previous approaches with radiative heat have focused primarily on phase-change materials~\cite{Abdallah_ThermalTransistor, Fan_RectivicationVacuumPRL2010, VO2NFHT_Novotny2016, Zwol_PhaseChangeVO22011, Zwol_PhaseChange2011}, electrostatically-tunable materials \cite{Papadakis_TunableNFHT}, and others~\cite{Boriskina_Ferroelectric2015, Khandekar_KerrPRB, Khandekar_FourwavemixingAPL}, here we carry out a general analysis on the basis of coupled-mode theory, and we identify the fundamental physical properties required for efficient switching of radiative heat. We conclude that low-loss and ultra-narrow band (i.e. high-quality factor) electromagnetic resonances are key for efficient switching of radiative heat. These results are to be contrasted with previous considerations based on the temperature-dependent emissivity of phase-change tunable materials where the contrast ratio is far smaller since the emissivity of the phase change materials have much broader bandwidth ~\cite{Abdallah_ThermalTransistor, Fan_RectivicationVacuumPRL2010}.}

\par{In search for a suitable material system for employing our concept of thermal switching, we leverage the low-loss and narrow-band response of monolayer hexagonal boron nitride (hBN) near the longitudinal optical (LO) phonon frequency \cite{Basov_hBN}. Significant progress has been recently made regarding the mechanical properties of hBN~\cite{AndroulidakisReview_2018,CastellanosReview2015,GraphenehBN_strain2015}. Recent findings have shown a strong sensitivity of the LO frequency of hBN with in-plane strain~\cite{Strain_hBN1}. Here, we derive the strain-dependent dielectric response of hBN entirely \emph{ab initio}, including the phonon-polariton frequency and linewidth. We find that the LO frequency redshifts by approximately $1$ meV per $0.1\%$ change in the lattice constant. Based on these findings, we propose a hBN-based photonic thermal switch that can achieve significant tuning of near-field radiative thermal conductance. We note that recently, strain-tunable near-field heat transfer with monolayer black phosphorus was considered \cite{BP_switch2020}. However, as we discuss below, the broad band characteristics of plasmons in black phosphorus are non-ideal for tuning radiative heat transfer. By contrast, here we show modulation of thermal conductance by more than an order of magnitude, enabled by the narrow band characteristics of the LO phonon of hBN.}

\section{\label{Concept}Theory of a three-resonator-based thermal switch}

  \begin{figure}
\centering
\includegraphics[width=0.8\linewidth]{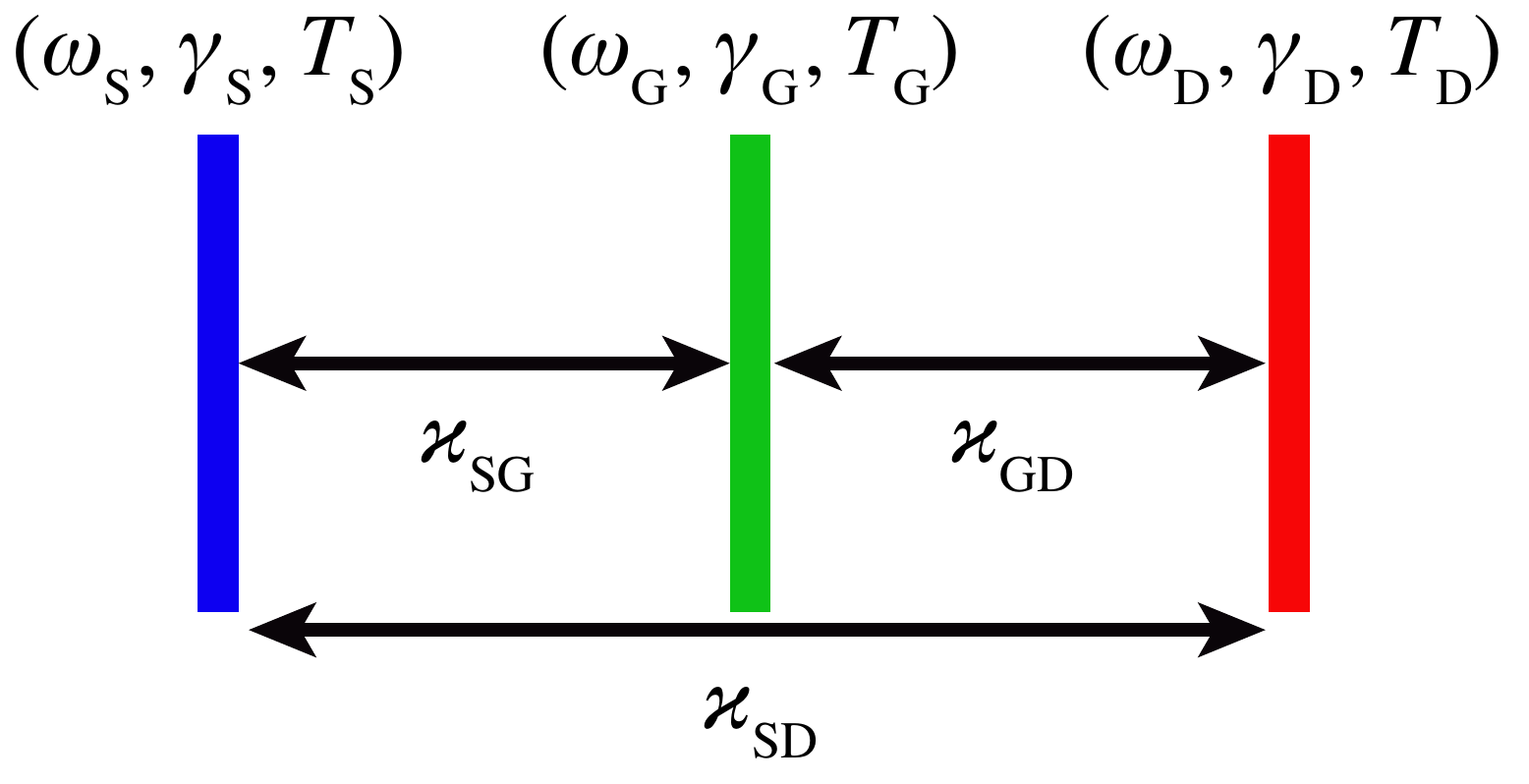}
\caption{Concept of a photonic thermal switch based on resonant coupling between thermally excited modes. Resonator $S$ represents the source, resonator $G$ the gate, and resonator $D$ the drain.}
\label{fig:Figure1} 
\end{figure}

\par{We illustrate the mechanism for high-contrast thermal switching by considering a theoretical model of three photonic resonators, as displayed in Fig.~\ref{fig:Figure1}. In parallel to the electronic transistor that functions by controlling the current that runs from the source to the drain via the gate, resonator $S$ represents the source, resonator $G$ represents the gate, and resonator $D$ is the drain, and we are interested in the thermal current received by the drain, $J_\mathrm{D}$. The gate is subject to a tuning mechanism of its physical property $X_\mathrm{G}$ that allows active modulation of $J_\mathrm{D}$. We note that previous work in~\cite{Abdallah_ThermalTransistor} considered a similar geometry for modulating thermal currents, however the principle of operation in~\cite{Abdallah_ThermalTransistor} is the phase-change of a gate composed of VO$_\mathrm{2}$, and does not depend on resonant coupling. Here, in contrast, we carry out an analytical theory of a three-resonator-based thermal switch and lay out general requirements for achieving large thermal tunability for \textit{any} physical property $X_\mathrm{G}$, as long as $X_\mathrm{G}$ manifests itself as a change in the electromagnetic modes supported in the gate.}

\par{In the system of Fig.~\ref{fig:Figure1}, resonators $S$, $G$, and $D$ are electromagnetically coupled. Hence, assuming that at each resonator alone there exists a localized electromagnetic mode with amplitude $\alpha_\mathrm{i}$, for $i=S, G, D$, this system can be described by coupled-mode theory (CMT)~\cite{Hauss}. Let us assume that each mode has an intrinsic decay rate described by $\gamma_\mathrm{i}$, and a resonant frequency $\omega_\mathrm{i}$, $i=S, G, D$. Via the fluctuation-dissipation theorem, assuming that each resonator is at a non-zero temperature $T_\mathrm{i}$, the intrinsic loss of each mode suggests that it is coupled to a compensating noise source, $n_\mathrm{i}$. Since the gate is subject to a tuning mechanism of its physical property $X_\mathrm{G}$, $\omega_\mathrm{G}$ depends on $X_\mathrm{G}$, and we write $\omega_\mathrm{G}=\omega_\mathrm{G}(X_\mathrm{G})$. The coupled-mode equations of this system are:
\begin{equation}\label{eq:1}
  \begin{aligned}
   d\alpha_\mathrm{S}/dt=i\omega_\mathrm{S}\alpha_\mathrm{S}-\gamma_\mathrm{S}\alpha_\mathrm{S}+2\sqrt{\gamma_\mathrm{S}}n_\mathrm{S}+i \kappa_\mathrm{SG}\alpha_\mathrm{G}+i \kappa_\mathrm{SD}\alpha_\mathrm{D} \\
   d\alpha_\mathrm{G}/dt=i\omega_\mathrm{G}(X_\mathrm{G})\alpha_\mathrm{G}-\gamma_\mathrm{G}\alpha_\mathrm{G}+2\sqrt{\gamma_\mathrm{G}}n_\mathrm{G}+i \kappa_\mathrm{SG}^{*}\alpha_\mathrm{S}+i \kappa_\mathrm{GD}\alpha_\mathrm{D} \\ 
   d\alpha_\mathrm{D}/dt=i\omega_\mathrm{D}\alpha_\mathrm{D}-\gamma_\mathrm{D}\alpha_\mathrm{D}+2\sqrt{\gamma_\mathrm{D}}n_\mathrm{D}+i \kappa_\mathrm{SD}^{*}\alpha_\mathrm{S}+i \kappa_\mathrm{GD}^{*}\alpha_\mathrm{G}
  \end{aligned}
  \end{equation}
where we normalize the amplitude of each mode so that $\left|\alpha_\mathrm{i}\right|^2$ expresses its energy. In Eq.~\ref{eq:1}, the parameters $\kappa_\mathrm{ij}$, for $i \neq j$ express the coupling rate between modes $i$ and $j$. The correlation of the noise sources is given by:
\begin{equation}\label{eq:2}
\langle{n_\mathrm{i}(\omega)n_\mathrm{i}^{*}(\omega^{'})}\rangle=\Theta(\omega, T_\mathrm{i})2\pi\delta(\omega-\omega^{'})
\end{equation}
for $i=S, G, D$. Here, $\Theta(\omega, T_\mathrm{i})$ is the mean energy per photon, given by:
\begin{equation}\label{eq:3}
\Theta(\omega, T_\mathrm{i})=\frac{\hbar\omega}{e^\mathrm{\hbar\omega/kT}-1}.
\end{equation}
Assuming harmonic fields ($\sim e^{i\omega t}$), the system of equations in Eq. ~\ref{eq:1} can be solved to compute $\alpha_\mathrm{i}(t)$, for $i=S, G, D$ (see supplemental information).}

\par{Due to the electromagnetic coupling between each pair of resonators, $J_\mathrm{D}$ has two components: one arising from its thermal exchange with the source, $J_{SD}$, and another from its thermal exchange with the gate, $J_{GD}$.  Let us assume that $T_\mathrm{S}>T_\mathrm{G}>T_\mathrm{D}$. In units of energy, we have:
\begin{equation}\label{eq:4}
J_\mathrm{D}=\int_{-\infty}^{\infty} J_\mathrm{D}(\omega) d\omega=\int_{-\infty}^{\infty} [J_\mathrm{SD}(\omega)+J_\mathrm{GD}(\omega)]d\omega
\end{equation}
where the spectral heat (which is not the Fourier transform of $J_\mathrm{D}$) flux $J_\mathrm{D}(\omega)$ consists of components:
\begin{equation}\label{eq:5}
   \begin{aligned}
  J_\mathrm{SD}(\omega)=\frac{1}{2\pi^2}\mathrm{Im}[\kappa_\mathrm{SD}\langle{\alpha_\mathrm{S}^{*}\alpha_\mathrm{D}}\rangle], \\  J_\mathrm{GD}(\omega)=\frac{1}{2\pi^2}\mathrm{Im}[\kappa_\mathrm{GD}\langle{\alpha_\mathrm{G}^{*}\alpha_\mathrm{D}}\rangle]. 
     \end{aligned}
\end{equation}
}

\par{In order to identify the conditions for high-contrast thermal switching of $J_\mathrm{D}$ via tuning the property $X_\mathrm{G}$, let us first examine a symmetric switch. In this case, the resonant frequencies of the modes in the source and drain are the same, i.e. $\omega_\mathrm{S}=\omega_\mathrm{D}$, and, furthermore, $\kappa_\mathrm{SG}=\kappa_\mathrm{GD}$. We further assume that $\kappa_\mathrm{SD}$ and $\kappa_\mathrm{GD}$ are small with respect to $\omega_\mathrm{i}$, for $i=S, G, D$.  (see supplemental information). By also considering the same loss rate for all modes, i.e. that $\gamma_\mathrm{S}=\gamma_\mathrm{G}=\gamma_\mathrm{D}=\gamma$, we can solve the system of equations in Eqs. \ref{eq:1}, \ref{eq:2} to obtain the spectral heat flux:
\begin{equation}\label{eq:6}
  \begin{aligned}
J_\mathrm{D}(\omega)=\frac{2\gamma^2}{\pi(\delta\omega_\mathrm{S}^2+\gamma^2)}\{ \kappa_\mathrm{SD}^2\frac{\Delta\Theta(\omega,T_\mathrm{S},T_\mathrm{D})}{\delta\omega_\mathrm{S}^2+\gamma^2} \\+\kappa_\mathrm{GD}^2\frac{\Delta\Theta(\omega,T_\mathrm{G},T_\mathrm{D})}{\delta\omega_\mathrm{G}^2+\gamma^2} 
  +\kappa_\mathrm{SD}\kappa_\mathrm{GD}^2\frac{(\omega_\mathrm{S}-\omega_\mathrm{G})\Theta(\omega,T_\mathrm{S})}{(\delta\omega_\mathrm{S}^2+\gamma^2)(\delta\omega_\mathrm{G}^2+\gamma^2)} \}
  \end{aligned}
  \end{equation}
where $\Delta\Theta(\omega,T_\mathrm{i},T_\mathrm{j})=\Theta(\omega,T_\mathrm{i})-\Theta(\omega,T_\mathrm{j})$, and $\delta\omega_\mathrm{i}=\omega-\omega_\mathrm{i}$, for $i,j=S, G, D$. By integrating Eq. \ref{eq:6} we obtain the total current $J_\mathrm{3}$, which is given by: 
 \begin{equation}\label{eq:6b}
   \begin{aligned}
J_\mathrm{D}=\frac{\kappa_\mathrm{SD}^2}{\gamma}\Delta\Theta(\omega_\mathrm{S},T_\mathrm{S},T_\mathrm{D}) \\ +\frac{2\kappa_\mathrm{GD}^2 \gamma}{(\omega_\mathrm{S}-\omega_\mathrm{G})^2+4\gamma^2}(\Delta\Theta(\omega_\mathrm{S},T_\mathrm{G},T_\mathrm{D})+\Delta\Theta(\omega_\mathrm{G},T_\mathrm{G},T_\mathrm{D})).
\end{aligned}
  \end{equation}}

\par{As can be seen by Eq. \ref{eq:6b}, $J_\mathrm{D}$ is maximum when $\omega_\mathrm{G}=\omega_\mathrm{S}$, i.e. when the mode of the gate is spectrally aligned with those of the source and drain. We refer to this state of maximum $J_\mathrm{D}$ as the ON state, where $J_\mathrm{D, ON}$ is given by:
  \begin{equation}\label{eq:7}
J_\mathrm{D, ON}=\frac{\kappa_\mathrm{SD}^2}{\gamma}\Delta\Theta(\omega_\mathrm{S},T_\mathrm{S},T_\mathrm{D})+\frac{\kappa_\mathrm{GD}^2}{\gamma}\Delta\Theta(\omega_\mathrm{S},T_\mathrm{G},T_\mathrm{D}).
\end{equation}
Eq. \ref{eq:7} clearly shows that the thermal current to the drain comprises of its thermal exchange with the source ($J_\mathrm{SD}$) and with that of the gate ($J_\mathrm{GD}$), weighted by their respective coupling constants, $\kappa_\mathrm{SD}$ and $\kappa_\mathrm{GD}$, respectively. Importantly, from Eq. \ref{eq:7} we see that in order to maximize $J_\mathrm{D, ON}$, which is desirable for the efficient operation of the switch, the loss factor $\gamma$ should be minimized.}

\par{As $\omega_\mathrm{G}$ deviates from $\omega_\mathrm{S}$, $J_\mathrm{D}$ reduces as can be seen from Eq. \ref{eq:6b}. In the OFF-state, we assume that $\omega_\mathrm{G}$ is maximally deviated from $\omega_\mathrm{S}$ as allowed by the tuning mechanism $X_\mathrm{G}$. We denote such an $\omega_\mathrm{G}(X_\mathrm{G,max})=\omega_\mathrm{G}^{\mathrm{o}}$. The current $J_\mathrm{D, OFF}$ at the OFF state can then be obtained using Eq. \ref{eq:6b} by setting $\omega_\mathrm{G}=\omega_\mathrm{G}^{\mathrm{o}}$. For high-contrast switching operation, the contrast ratio between the ON and OFF states, defined as $(J_\mathrm{D, ON}-J_\mathrm{D, OFF})/J_\mathrm{D, ON}$, should be large. With Eqs. \ref{eq:6b}, \ref{eq:7}, we obtain the contrast ratio:
\begin{equation}\label{eq:8}
\kappa_\mathrm{GD}^2\frac{\Delta\Theta(\omega_\mathrm{S},T_\mathrm{G},T_\mathrm{D})-\alpha(\Delta\Theta(\omega_\mathrm{S},T_\mathrm{G},T_\mathrm{D})+\Delta\Theta(\omega_\mathrm{G}^{\mathrm{o}},T_\mathrm{G},T_\mathrm{D}))}{\kappa_\mathrm{SD}^2\Delta\Theta(\omega_\mathrm{S},T_\mathrm{S},T_\mathrm{D})+\kappa_\mathrm{GD}^2\Delta\Theta(\omega_\mathrm{S},T_\mathrm{G},T_\mathrm{D})},
\end{equation}
where $\alpha$ is given by 
  \begin{equation}\label{eq:9}
\alpha=\frac{2\gamma^2}{(\omega_\mathrm{S}-\omega_\mathrm{G}^{\mathrm{o}})^2+4\gamma^2}.
\end{equation}
}

\par{From this analysis we can identify the conditions for an efficient thermal switch: (i) as stated above, enhancing $J_\mathrm{D, ON}$ requires a small loss rate $\gamma$, wheres (ii) maximizing thermal modulation requires minimizing the parameter $\alpha$ in Eq. \ref{eq:9}, which in turn suggests that  $\omega_\mathrm{G}^{\mathrm{o}}$ ought to differ significantly from  $\omega_\mathrm{S}$. In the limit of $\gamma\rightarrow0$, it can be seen that infinitesimal deviation of $\omega_\mathrm{G}^{\mathrm{o}}$ from $\omega_\mathrm{S}$ suffices for efficient thermal switching. Therefore, we have shown here that an ultra-low-loss system composed of three resonators can serve as an efficient thermal switch, provided that an intrinsic property of the gate ($X_\mathrm{G}$) can be tuned sufficiently. By operating near a photonic resonance of the system ($\omega_\mathrm{G}\sim\omega_\mathrm{S}$), even small changes in the property $X_\mathrm{G}$ can yield significant tuning of the thermal current $J_\mathrm{D}$. In what follows, we show that such requirements are satisfied by monolayer hBN, via tuning its dielectric properties with in-plane strain.}
 
\section{\label{Abinitio}First-principles calculation of hBN optical conductivity}

\begin{figure}
\centering
\includegraphics[width=1\linewidth]{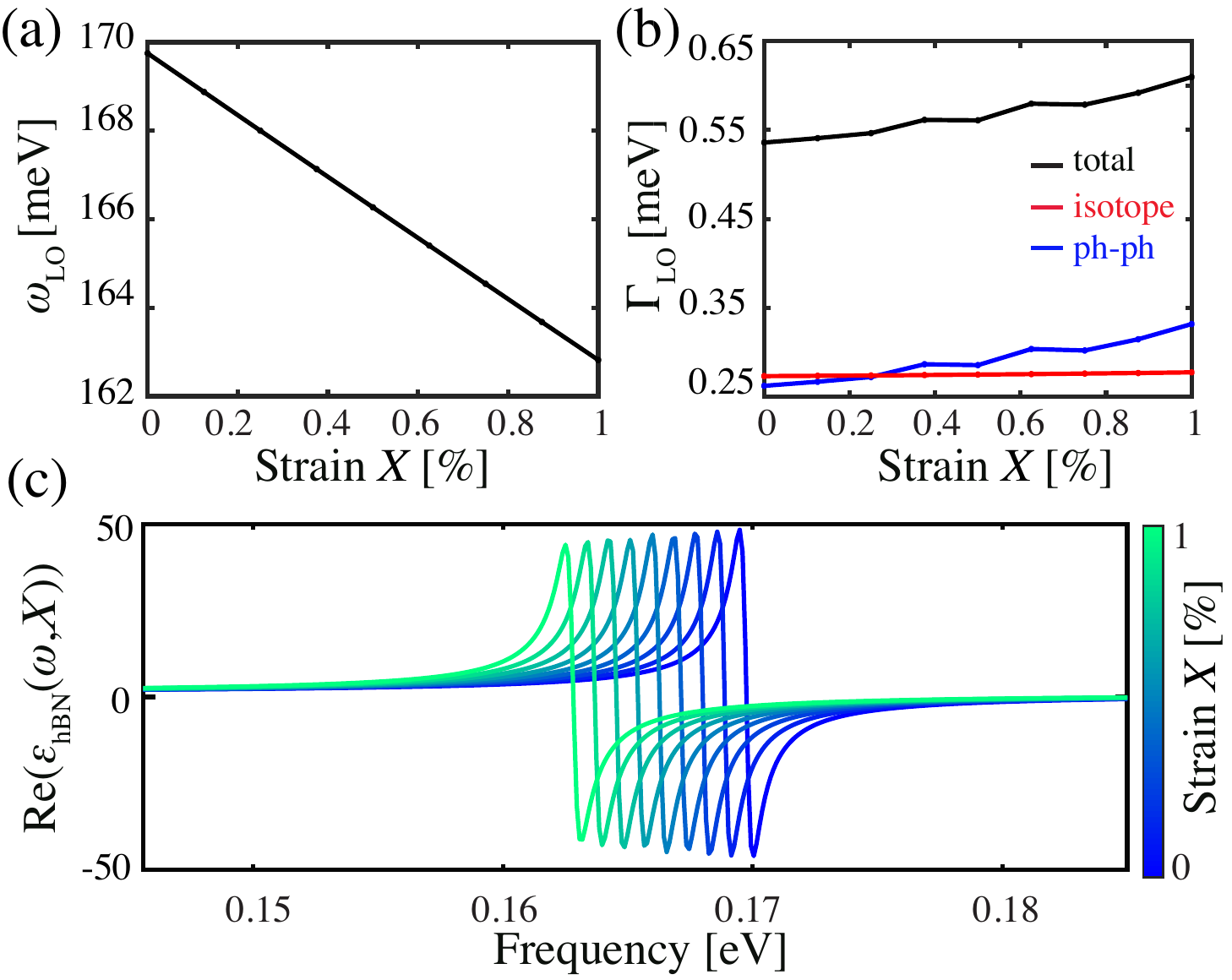}
\caption{(a) Resonance frequency $\omega_\mathrm{LO}$ and (b) linewidth, $\Gamma_\mathrm{LO}$, of the longitudinal optical phonon of monolayer hBN as a function of strain, computed ab initio (\cite{RiverahBN2019}, see discussion around Eq.~\ref{eq:2Dcond}). (c) Real part of the dielectric function of monolayer hBN, $\epsilon_\mathrm{hBN}(\omega,X)$, as a function of frequency for different strains, $X$.}
\label{fig:Figure2} 
\end{figure}

\par{In the previous section we showed that a fundamental requirement for an efficient thermal switch is a low-loss electromagnetic resonance. In search of a material system that supports such low-loss photonic resonances, we consider that the plasmonic, excitonic, and phononic resonances of numerous materials become pronounced in their monolayer form~\cite{AbajoPolaritons2016, Papadakis_polaritons}. In particular, monolayer hBN supports a low-loss LO phonon in the mid-infrared range, namely near $170$ meV ~\cite{Basov_hBN, RiverahBN2019}, which corresponds to the peak of moderate-temperature thermal emission. When coupled to radiation, the resulting surface phonon polariton (SPhP) is characterized by a very large quality factor~\cite{Cadwell_Phonons, hBN_Jabob}. Furthermore, along with other van der Waals material systems, hBN exhibits excellent mechanical properties and can sustain large in-plane deformations without breaking~\cite{AndroulidakisReview_2018,CastellanosReview2015,GraphenehBN_strain2015}. It was recently experimentally demonstrated that small changes in the in-plane lattice constant of monolayer hBN, on the order of $0.2\%$, can yield a significant shift of its LO phonon frequency~\cite{Strain_hBN1}, on the order of few tens of meV's. The low-loss and strain-tunable nature of the SPhP mode in hBN represent ideal characteristics for the implementation of the photonic thermal switch discussed above.}

\par{In this section, we carry out first principles calculations of the dielectric function of monolayer hBN, and estimate the strain-induced shifting of both its phonon frequency, $\omega_\mathrm{LO}$, as well as its linewidth, $\Gamma_\mathrm{LO}$. The atomic motion associated with the LO phonon mode of hBN introduces a polarization density and corresponding electric field~\cite{RiverahBN2019}. This can be related to the 2D optical conductivity, which has the general form~\cite{RiverahBN2019}: 
\begin{equation}
\sigma(\mathbf{q},\omega) = - \frac{i \omega}{\Omega} 
\frac{\left| \hat{\mathbf{q}} \cdot \sum_\kappa \mathbf{Z}_\kappa \boldsymbol{\eta}_\kappa\right|^2}
{\omega_{\mathrm{\textbf{q},LO}}^2 - \omega^2 - i \omega \Gamma_{\mathrm{\textbf{q},LO}}}.
\label{eq:2Dcond}
\end{equation}
Here, $\mathbf{q}$ is the wavevector, $\Omega$ is the unit cell area of the hBN flake, $\mathbf{Z}_\kappa$ is the Born effective charge tensor for atom $\kappa$ in the unit cell and $\boldsymbol{\eta}_\kappa$ is the corresponding mass-weighted eigendisplacement of the LO phonon mode. Overall, the optical conductivity is wavevector-dependent, however, here we only consider response from the zone center, i.e., the $\Gamma$ point of the Brillouin zone, where $\mathbf{q} \rightarrow 0$.}

\par{All quantities in Eq.~\ref{eq:2Dcond} are computed using first principles calculations. The phonon linewidth is determined by considering scattering of the LO phonon mode with other phonons and isotopic impurities. In principle, all of these phonon properties vary with temperature, for instance due to changes in the lattice constant upon heating. In this work, we only consider temperature effects in the phonon linewidths, where scattering events are weighted by temperature-dependent Bose-Einstein occupation factors. In Eq.~\ref{eq:2Dcond} we set the linewidth $\Gamma_\mathrm{LO}$ to twice the imaginary part of the phonon self-energy, computed \emph{ab initio}.}

\par{We use the formalism outlined above to model the optical conductivity (Eq.~\ref{eq:2Dcond}) as a function of biaxial, in-plane strain of monolayer hBN. In our work, we consider tensile strain relative to the relaxed lattice constant. In Fig.~\ref{fig:Figure2}a, we show the evolution of the LO phonon frequency as a function of tensile strain of up to 1\%. Similar to previous work~\cite{Strain_hBN1,li_thermal_2017}, we find that the phonon frequency linearly redshifts with increasing tensile strain, with a slope of $-6.8$ meV/\%, in agreement with experiments \cite{Strain_hBN1}. This redshift is anticipated based on the increase of the phonon oscillation amplitude with strain. In Fig. \ref{fig:Figure2}b, we show the linewidth of the LO phonon that is the sum two terms, arising from isotope scattering and the phonon-phonon scattering (see supplemental information).} 

\par{Based on the computed conductivity of Eq. \ref{eq:2Dcond}, we estimate the dielectric function of hBN via $\epsilon_\mathrm{hBN}=1+i\sigma_\mathrm{hBN}(\omega)/(\epsilon_\mathrm{o} \omega t)$~\cite{Falkovsky}, where $\epsilon_\mathrm{o}$ is the dielectric constant of free-space, and $t$ is the thickness of the hBN monolayer, taken as $t=0.317$ nm~\cite{hBN_thickness}. As can be seen from Fig.~\ref{fig:Figure2}c, the LO phonon results in a Lorentzian-shaped dielectric function that is ultra-narrow band and highly tunable with strain. In what follows, we levarage the low-loss LO resonance and extreme tunability of hBN monolayers with strain to propose an efficient and deep subwavelength thermal switch.}

\section{\label{Results}Results}

\par
\begin{figure}
\centering
\includegraphics[width=1\linewidth]{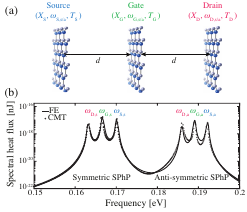}
\caption{(a) Schematic of the hBN-based thermal switch composed of three hBN monolayers at different strains $X_\mathrm{S}$, $X_\mathrm{G}$, and $X_\mathrm{D}$. (b) Spectral heat flux to the drain ($J_\mathrm{D}(\omega)$) for $X_\mathrm{S}=0\%$, $X_\mathrm{G}=0.5\%$, $X_\mathrm{D}=1\%$, calculated from coupled- mode theory (circles) and fluctuational electrodynamics (solid curve). The triplet of resonances near 0.17eV and near 0.1925 eV correspond to the symmetric and anti-symmetric SPhP modes, respectively, that dominate the near-field heat transfer. The temperatures are set to: $T_\mathrm{S}=500$ K, $T_\mathrm{D}=300$ K, $T_\mathrm{G}=430$ K.}
\label{fig:Figure3} 
\end{figure}

{In the previous section we demonstrated the significant redshift of the LO phonon frequency and associated LO linewidth increase with in-plane strain in monolayer hBN (Fig. \ref{fig:Figure2}). We consider a three-resonator-based thermal switch composed of hBN monolayers as depicted in Fig.~\ref{fig:Figure3}a. We consider a separation distance between hBN monolayers of $d=20$ nm, which is far smaller than the relevant thermal wavelength. This suggests that the surface-confined SPhP modes supported at each hBN monolayer are electromagnetically coupled and can thus be described by the formalism of Section~\ref{Concept}. Here, the physical property that we tune, termed $X_\mathrm{G}$ in Section~\ref{Concept}, is the strain of the gate-hBN. The strain in the source and drain, $X_\mathrm{S}$ and $X_\mathrm{D}$ respectively, remain fixed.}

\par{In the thermal switch of Fig.~\ref{fig:Figure3}a, the gate-hBN serves as an ultra-narrow-band filter of thermal radiation, which controls the passage of a thermal current to the drain. In parallel with the discussion in Section~\ref{Concept}, the resonant frequency of the SPhP mode of the gate ($\omega_\mathrm{G}$) is tunable with strain. We note, however, that in the structure of Fig.~\ref{fig:Figure3}a, there exist not only three modes, but in fact six, because each monolayer supports two modes: a symmetric and an anti-symmetric SPhP mode. These six modes are described by the properties ($\omega_\mathrm{S,s}$, $\gamma_\mathrm{S,s}$), ($\omega_\mathrm{G,s}$, $\gamma_\mathrm{G,s}$), ($\omega_\mathrm{D,s}$, $\gamma_\mathrm{D,s}$) and ($\omega_\mathrm{S,a}$, $\gamma_\mathrm{S,a}$), ($\omega_\mathrm{G,a}$, $\gamma_\mathrm{G,a}$), ($\omega_\mathrm{D,a}$, $\gamma_\mathrm{D,a}$), respectively. Due to the low-loss nature of the SPhP modes in monolayer hBN, as well as the fact that the asymmetric SPhP mode occurs at frequencies significantly higher than the symmetric SPhP mode, we do not consider here the coupling between symmetric and anti-symmetric modes. Furthermore, we take the temperatures of the source and drain to be $T_\mathrm{S}=500$ K and $T_\mathrm{D}=300$ K, respectively.}

\par{As discussed in Section~\ref{Concept}, the thermal current to the drain consists of two components, its thermal exchange with the source, and its thermal exchange with the gate, $J_\mathrm{D}=J_\mathrm{SD}+J_\mathrm{GD}$ (see Eqs.~\ref{eq:5}, \ref{eq:6}, \ref{eq:7}). Following \cite{PolderNFHT}, these thermal power exchanges can be computed with fluctuational electrodynamics by determining the ensemble average of the Poynting flux to the drain, while accounting for all thermally excited fluctuations, namely those of the source, gate, and drain. The thermal exchange between the source and the drain is described by:
\begin{equation}\label{eq:JSD}
J_\mathrm{SD}=\frac{1}{4\pi^2}\int_{-\infty}^{\infty} \Phi_\mathrm{SD}(\omega) [\Theta(\omega,T_\mathrm{S})-\Theta(\omega,T_\mathrm{D})] d\omega
\end{equation}
where $\Phi_\mathrm{SD}(\omega)$ is given by:
\begin{equation}\label{eq:PhiSD}
\Phi_\mathrm{SD}(\omega)=\int_{0}^{\infty}\xi_\mathrm{SD}(\omega,\beta)\beta d\beta.
\end{equation}
In Eq.~\ref{eq:PhiSD}, $\xi_\mathrm{SD}(\omega,\beta)$ is the probability of a photon to be transmitted between the source and the drain, in the presence of the gate, in the setup of Fig.~\ref{fig:Figure3}a, and $\beta$ is the in-plane wavenumber. Similarly, by exchanging the subscript $S\rightarrow G$ we can obtain the thermal exchange between the gate and the drain, $J_\mathrm{GD}$, in which case $\xi_\mathrm{GD}(\omega,\beta)$ is the probability of a photon to be transmitted between the gate and the drain, in the presence of the source. The photon transmission probabilities $\xi_\mathrm{SD}(\omega,\beta)$ and $\xi_\mathrm{GD}(\omega,\beta)$ can be computed via Fresnel's coefficients as introduced in~\cite{PolderNFHT} (see computational package in~\cite{CHEN2018163}).}

\par
\begin{figure}
\centering
\includegraphics[width=1\linewidth]{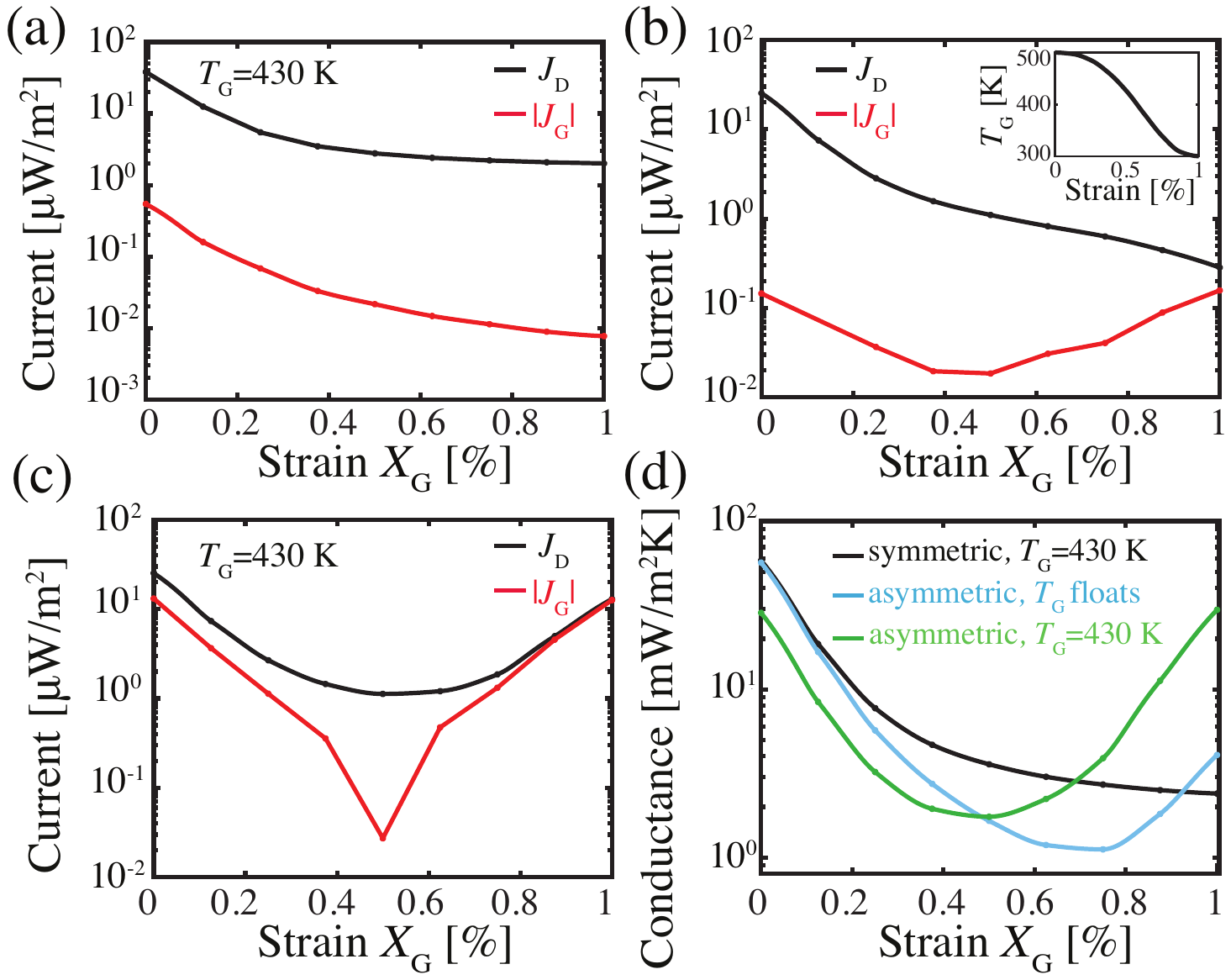}
\caption{Thermal currents $J_\mathrm{D}$ and $J_\mathrm{G}$ as a function of the strain in the gate, $X_\mathrm{G}$, for (a) the symmetric structure for $T_\mathrm{G}=430$ K, which minimizes $J_\mathrm{G}$, for (b) the asymmetric structure for a floating $T_\mathrm{G}$ (inset) that minimizes $J_\mathrm{G}$, and for (c) the asymmetric structure for $T_\mathrm{G}=430$ K. (d) Thermal conductance $Q$ as a function of the strain in the gate for the cases in panels (a), (b), (c), near $T=430$ K, as $T$ floats such that $T=T_\mathrm{G}$, and near $T=430$ K, respectively.}
\label{fig:Figure4} 
\end{figure}

\begin{figure}
\centering
\includegraphics[width=1\linewidth]{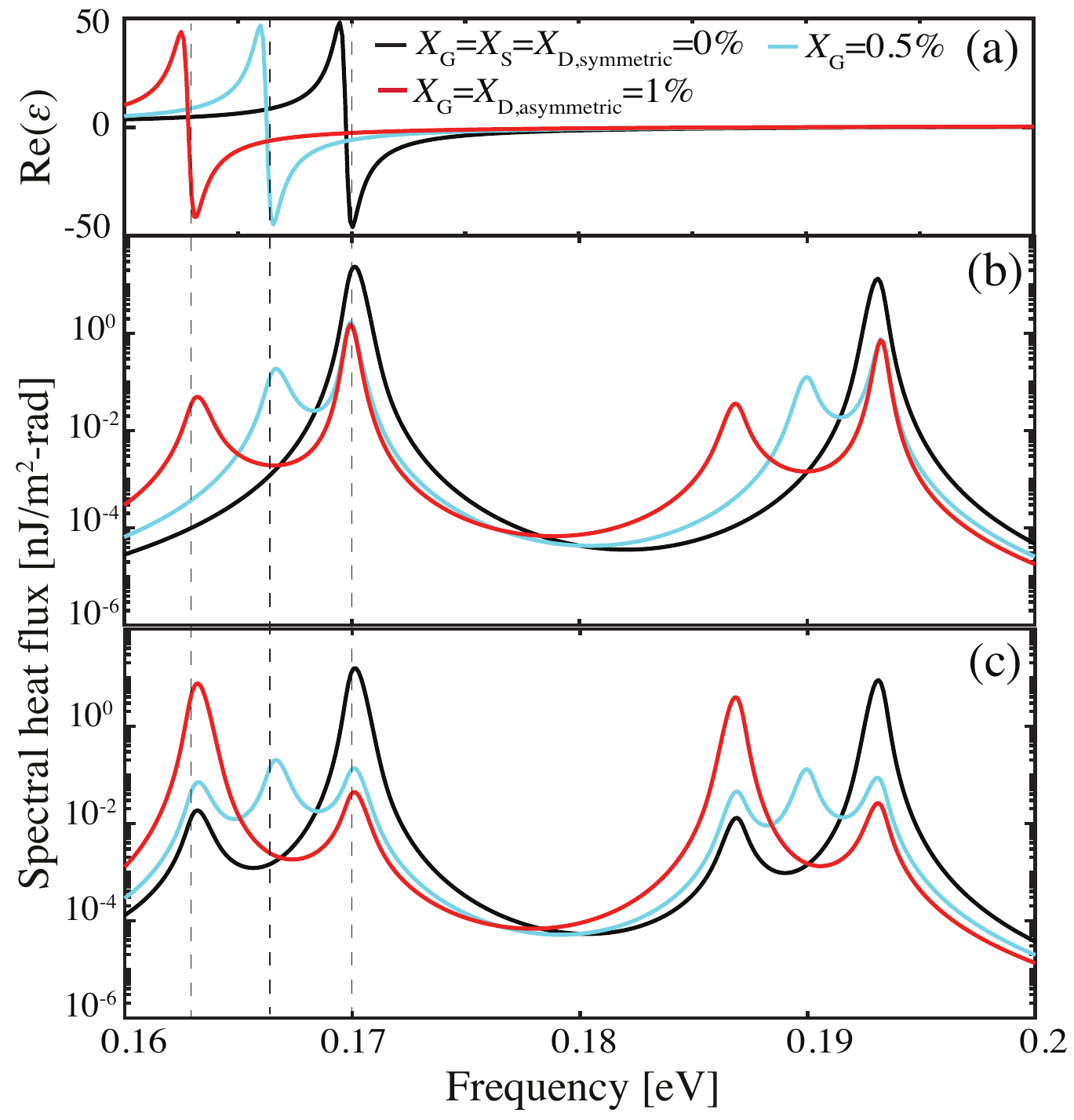}
\caption{(a) Real part of dielectric function of monolayer hBN for different strains. (b), (c) Spectral heat flux to the drain ($J_\mathrm{D}(\omega)$) for the symmetric and asymmetric thermal switches, respectively, for $T_\mathrm{S}=500$ K, $T_\mathrm{D}=300$ K, and $T_\mathrm{G}=430$ K. Black curves pertain to $X_\mathrm{G}=0\%$, cyan curves pertain to $X_\mathrm{G}=0.5\%$, and red curves pertain to $X_\mathrm{G}=1\%$.}
\label{fig:Figure5} 
\end{figure}

\begin{figure}
\centering
\includegraphics[width=1\linewidth]{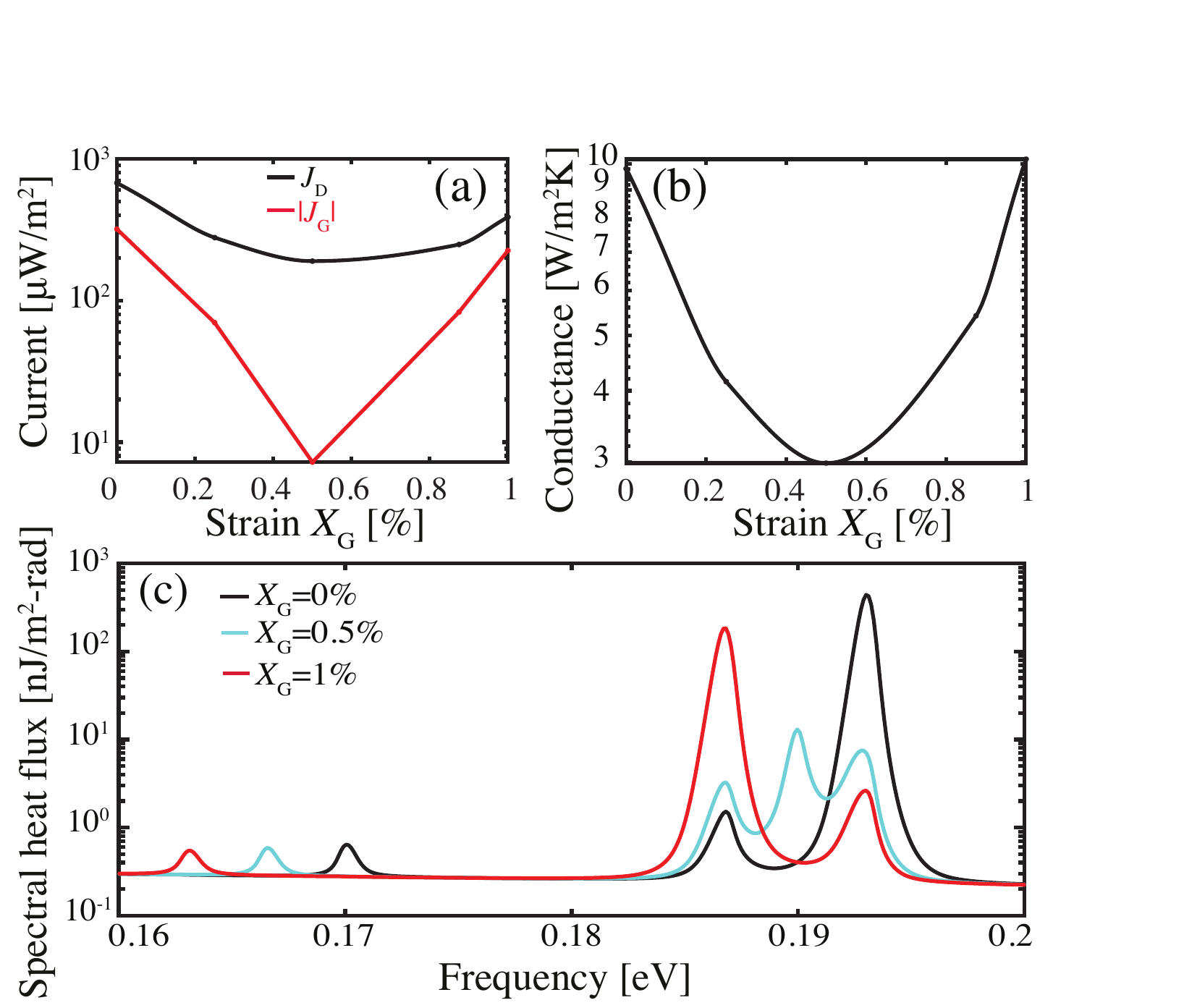}
\caption{(a) Thermal currents $J_\mathrm{D}$ and $J_\mathrm{G}$ as a function of the strain in the gate, for the photonic thermal switch with tungsten substrates on the source and the drain. (b) Corresponding thermal conductance $Q$ near $T=2700$ K. (c) Spectral heat flux to the drain ($J_\mathrm{D}(\omega)$), for $T_\mathrm{S}=3000$ K and $T_\mathrm{D}=2500$ K. Black curve pertains to $X_\mathrm{G}=0\%$, cyan curve pertains to $X_\mathrm{G}=0.5\%$, and red curve pertains to $X_\mathrm{G}=1\%$.}
\label{fig:Figure6} 
\end{figure}

{In evaluating the CMT formalism  developed in Section~\ref{Concept} with respect to fluctuational electrodynamics, let us first consider that heat exchange occurs at a single in-plane wavenumber ($\beta$) channel, namely for $\beta=2/d$~\cite{Pendry_NFHT1999}. As noted in Refs. \cite{Fan_RectivicationVacuumPRL2010,Pendry_NFHT1999}, this channel provides the largest contributions to the near-field heat transfer. We assume $X_\mathrm{S}=0\%$, $X_\mathrm{G}=0.5\%$, $X_\mathrm{D}=1\%$, so that the resonant frequencies of the SPhP modes supported at each hBN monolayer are easy to distinguish from one another in Fig. \ref{fig:Figure3}b. To ensure this, we also reduce the phonon linewidth of hBN by a factor of $2$ with respect to the results of Fig. \ref{fig:Figure2}b, for the sake of clearly demonstrating the interaction of the SPhP modes in this system (We note that for the numerical results in Figs. \ref{fig:Figure4}-\ref{fig:Figure6} we use the $\Gamma_\mathrm{LO}$ shown in Fig. \ref{fig:Figure2}b). These choices yield two distinct triplets of peaks in the spectral heat exchange shown in Fig.~\ref{fig:Figure3}b, where the CMT result is shown with the circles, while the result with fluctuational electrodynamics is plotted with the solid curve. The triplet of peaks centered around $0.165$ eV arises from the excitation of the symmetric SPhP mode at each hBN monolayer, whereas the triplet of peaks near $0.19$ eV corresponds to the anti-symmetric SPhP modes. For each triplet of peaks, the high-frequency one corresponds to the source ($X_\mathrm{S}=0\%$), the intermediate peak corresponds to the gate ($X_\mathrm{G}=0.5\%$), and the low-frequency one corresponds to the drain ($X_\mathrm{D}=1\%$), as expected since the dielectric function of hBN redshifts as strain increases (Fig.~\ref{fig:Figure2}a).} 

\par{The parameters $\omega_\mathrm{S,s/a}$, $\omega_\mathrm{G,s/a}$, $\omega_\mathrm{D,s/a}$, representing the symmetric/anti-symmetric resonant frequencies of the SPhP modes in the source, gate, and drain, respectively, as shown in Fig. \ref{fig:Figure3}b, are initially estimated from the single-slab SPhP dispersion relation via the reflection-pole method~\cite{Anemogiannis_RPM}. Furthermore, the upper bound of the loss rate of an electromagnetic mode supported in a single-mode nanostructure is half of its material loss~\cite{Raman_upperboundbeta_2013}. Therefore, since these hBN monolayers constitute extremely low-loss systems, we take the loss rates of the modes to be $\gamma_\mathrm{S,s/a}=\Gamma_\mathrm{LO,S}/2$, $\gamma_\mathrm{G,s/a}=\Gamma_\mathrm{LO,G}/2$, $\gamma_\mathrm{D,s/a}=\Gamma_\mathrm{LO,D}/2$, where $\Gamma_\mathrm{LO,S/G/D}$ is the LO phonon linewidth of hBN at the source, gate, and drain, respectively, normalized here by a factor of $2$ for the sake of clearly demonstrating the physics as mentioned above. The coupling rates ($\kappa_\mathrm{SD}$, $\kappa_\mathrm{GD}$, $\kappa_\mathrm{SG}$) as well as final values of the modes resonant frequencies and loss rates are determined via curve-fitting of the CMT model to the result with fluctuational electrodynamics. These are provided in the supplemental information. As can be seen in Fig.~\ref{fig:Figure3}b, the CMT model is in good agreement with fluctuational electrodynamics, suggesting that the set of equations in Eq. ~\ref{eq:1} that describes the system of Fig.~\ref{fig:Figure3}a as a triplet of resonators is indeed appropriate.}

\par{Now that we established the validity of the CMT formalism introduced in Section~\ref{Concept}, we proceed in evaluating the performance of the hBN thermal switch in terms of tunability of the thermal current to the drain, $J_\mathrm{D}$. We will consider two cases: a ``symmetric'' case, for which $X_\mathrm{S}=X_\mathrm{D}=0\%$, and an ``asymmetric'' case for which $X_\mathrm{S}=0\%$ while $X_\mathrm{D}=1\%$. The strain in the gate is tuned between $0\%\leq X_\mathrm{G} \leq 1\%$.}

\par{An efficient thermal switch ought to operate with minimum current passing through its gate ~\cite{Li_ThermalTransistor, Abdallah_ThermalTransistor,Three_bodies_Abdallah2012}. This current is given by $J_\mathrm{G}=J_\mathrm{GS}+J_\mathrm{GD}$, where $J_\mathrm{GS}$ and $J_\mathrm{GD}$ can be obtained by the appropriate interchange of indices in Eqs.~\ref{eq:JSD}, \ref{eq:PhiSD}. The condition $J_\mathrm{G}=0$ determines the temperature of the gate, $T_\mathrm{G}$, in the results that follow. Due to the very narrow bandwidth of the dielectric resonance of hBN near its LO phonon (Figs. \ref{fig:Figure2}b, c), we evaluate the integrals of Eq.~\ref{eq:JSD} at $\omega_\mathrm{G}=\omega_\mathrm{LO,G}$ (as shown in Fig.~\ref{fig:Figure2}a). Hence, the condition $J_\mathrm{G}=0$ is evaluated by: 
\begin{equation}\label{eq:TG}
n(\omega_\mathrm{G},T_\mathrm{G})=\frac{n(\omega_\mathrm{G},T_\mathrm{S})\Phi_\mathrm{GS}(\omega_\mathrm{G})+n(\omega_\mathrm{G},T_\mathrm{D})\Phi_\mathrm{GD}(\omega_\mathrm{G})}{\Phi_\mathrm{GS}(\omega_\mathrm{G})+\Phi_\mathrm{GD}(\omega_\mathrm{G})},
\end{equation}
where $n(\omega,T)=(e^{\hbar \omega/k T}-1)^{-1}$ is the photon occupation number.}

\par{Eq.~\ref{eq:TG} determines the temperature of the gate for minimizing $J_\mathrm{G}$. In the symmetric switch, $\Phi_\mathrm{GS}=\Phi_\mathrm{GD}$ due to symmetry with respect to the plane of the gate-hBN (see Fig.~\ref{fig:Figure3}a), therefore the condition in Eq.~\ref{eq:TG} reduces to $n(\omega_\mathrm{G},T_\mathrm{G})=[n(\omega_\mathrm{G},T_\mathrm{S})+n(\omega_\mathrm{G},T_\mathrm{D})]/2$. In this case, $T_\mathrm{G}$ remains roughly constant at $T_\mathrm{G}=430$ K for all levels of strain in the gate. We show in Fig.~\ref{fig:Figure4}a the thermal currents $J_\mathrm{D}$ and $J_\mathrm{G}$, where it can be seen that $J_\mathrm{D}$ is tuned by more than an order of magnitude as the gate-hBN is strained, while $J_\mathrm{G}$ remains two orders of magnitude smaller than $J_\mathrm{D}$.}

\par{As expected, maximum heat transfer occurs when the SPhP frequencies in the source, gate, and drain are spectrally overlapping, i.e. when the three hBN monolayers are at the same level of strain. Since $X_\mathrm{S}=X_\mathrm{D}=0\%$, this is achieved when $X_\mathrm{G}$ is also $0\%$. The decrease in $J_\mathrm{D}$ with strain that is shown in Fig.~\ref{fig:Figure4}a can be explained by considering Eq. \ref{eq:6b}, where it was shown that as $\omega_\mathrm{2}$ deviates from $\omega_\mathrm{1}$, in other words as the gate hBN monolayer is strained, the heat transfer decreases. As shown from Fig. \ref{fig:Figure4}a, indeed, as $X_\mathrm{G}$ increases, the thermal current to the gate decreases as a result of the spectral mismatch between the resonance frequency of the source and drain hBN with respect to that of the gate.}

\par{These findings can also be seen in the spectral heat flux $J_\mathrm{D}(\omega)$ of Fig.~\ref{fig:Figure5}b. At $X_\mathrm{G}=0$ (black curve), the two distinct peaks near $0.17$ eV and near $0.195$ eV correspond to the symmetric and anti-symmetric modes that are supported in the source, gate, and drain, simultaneously. The spectral position of the symmetric mode is aligned with the condition $\epsilon_\mathrm{hBN}(\omega_\mathrm{LO},0\%)=-1$, as shown with the black curve in Fig.~\ref{fig:Figure5}a. As $X_\mathrm{G}$ increases to $0.5\%$, the dielectric function of the gate-hBN redshifts (cyan curve in Fig.~\ref{fig:Figure5}a), and the spectral heat flux exhibits an additional set of symmetric and anti-symmetric peaks as shown in Fig.~\ref{fig:Figure5}b (cyan curve). The amplitude of this spectrum is smaller than in the case of $X_\mathrm{G}=0\%$, as expected due to the spectral misalignment between the SPhP of the gate with respect to those of the source and drain. For $X_\mathrm{G}=1\%$, the dielectric function of the gate-hBN redshifts even more, as shown with the red curve in Fig.~\ref{fig:Figure5}a. Hence, the amplitude of the spectral heat flux shown with the red curve in Fig.~\ref{fig:Figure5}b reduces even more, as a result of the spectral mismatch between the SPhP of the gate with those of the source and drain.}

\par{In Fig.~\ref{fig:Figure4}d, we plot the thermal conductance , defined as: 
\begin{equation}\label{eq:conductance}
Q(T)=\frac{1}{4\pi^2}\int_{-\infty}^{\infty} \Phi(\omega) \frac{\partial \Theta(\omega,T)}{\partial T} d\omega.
\end{equation}
where $\Phi(\omega)=\Phi_\mathrm{SD}(\omega)+\Phi_\mathrm{GD}(\omega)$. With the black curve we show the thermal conductance of the symmetric structure discussed in panel (a), for $T$ near $340$ K, the temperature of the gate. As can be seen, the thermal conductance is tunable by more than an order of magnitude as the gate-hBN is strained from $X_\mathrm{G}=0\%$ to $X_\mathrm{G}=1\%$. The corresponding ON/OFF contrast ratio is $96\%$, therefore this thermal switch can significantly suppress and enhance heat transfer preferentially via strain.}

\par{Panels (b) and (c) in Fig.~\ref{fig:Figure4} pertain to the asymmetric structure for which $X_\mathrm{S}=0\%$ while $X_\mathrm{D}=1\%$. In panel (b) the temperature of the gate is selected such that the current through the gate is minimized, whereas in panel (c) $T_\mathrm{G}=430$ K. In the inset of Fig.~\ref{fig:Figure4}b we show the dependence of $T_\mathrm{G}$ on the strain in the gate for minimizing $J_\mathrm{G}$ (Eq. \ref{eq:TG}). The decrease in $T_\mathrm{G}$ with strain can be understood as follows: As mentioned previously, $J_\mathrm{G}=J_\mathrm{GS}+J_\mathrm{GD}$. As the strain in the gate increases, $\xi_\mathrm{GS}$ decreases since the strain in the gate deviates from that of the source ($X_\mathrm{S}=0\%$). However, as $X_\mathrm{G}$ increases,  $\xi_\mathrm{GD}$ increases because the strain in the gate approaches that of the drain ($X_\mathrm{D}=1\%$), therefore the resonant frequencies of the gate and drain align. To maintain $J_\mathrm{GD}$ small, therefore, as the strain in the gate increases, its temperature must approach that of the drain (see Eq.~\ref{eq:JSD} for $S\rightarrow G$).}

\par{As can be seen in Fig. \ref{fig:Figure4}b, $J_\mathrm{D}$ is tunable by more than two orders of magnitude, while $J_\mathrm{G}$ remains significantly below $J_\mathrm{D}$ for all levels of strain in the gate. In panel (d) we show with the cyan color the thermal conductance for $T$ near $T_\mathrm{G}$ as shown in the inset of panel (b). In this case, the thermal conductance can be tuned from $57$ mW/m$^{2}$K to $1$ mW/m$^{2}$K. This corresponds to an ON/OFF contrast ratio of $98\%$.}

\par{In Fig.~\ref{fig:Figure4}c, we show how the asymmetric thermal switch operates when $T_\mathrm{G}$ is fixed at $T_\mathrm{G}=430$ K. In this case, we obtain a thermal current $J_\mathrm{D}(X_\mathrm{G})$ that is symmetric with respect to $X_\mathrm{G}=0.5\%$. This is expected since at $X_\mathrm{G}=0\%$ the SPhP mode of the gate is spectrally aligned with that of the source, as shown with the black curve in the spectral heat flux of Fig.~\ref{fig:Figure5}c, whereas at $X_\mathrm{G}=1\%$ it are aligned with the SPhP mode of the drain, as shown with the red curve in Fig.~\ref{fig:Figure5}c. At $X_\mathrm{G}=0.5\%$, heat flux is at its minimum because the SPhP mode of the gate does not align with either the source or the drain. This leads to three peaks in the spectral heat flux of Fig.~\ref{fig:Figure5}c (cyan curve) for the symmetric SPhP modes: one for the drain, one for the gate, and one for the source, and similarly three peaks for the anti-symmetric SPhP modes. The thermal conductance of the asymmetric structure near $T=430$ K is shown in Fig.~\ref{fig:Figure4}d with the green curve. Similar to the thermal current $J_\mathrm{D}$, the thermal conductance is symmetric with respect to $X_\mathrm{G}=0.5\%$. We have shown in Fig.~\ref{fig:Figure4} that by controlling the strain in the gate, as well as the temperature of the gate, we can obtain qualitatively different responses, namely an enhancement or suppression of the heat conductance.}

\par{Finally, in Fig.~\ref{fig:Figure6} we consider a scenario where the source-hBN and the drain-hBN monolayers are placed on a thermally emitting substrate material. Our aim is to precisely control the amount of heat transferred from the source-substrate to the drain-substrate through the gate. As a substrate material, we select tungsten (W) because its emission resembles that of a black body, i.e. it is broad band. The optical properties of tungsten were obtained from~\cite{Palik_Book}. In order to explore how the proposed thermal switch operates at higher temperatures, we consider $T_\mathrm{S}=3000$ K and $T_\mathrm{D}=2500$ K, both of which remain below the melting point of W (nearly $3700$ K) and hBN ($3273$ K)~\cite{hBN_meltingpoint}. By letting $T_\mathrm{G}$ float such that it minimizes $J_\mathrm{G}$ (Eq.~\ref{eq:TG}), we obtain $T_\mathrm{G}\approx2700$ K. In Fig.~\ref{fig:Figure6}a, we show the thermal currents $J_\mathrm{D}$ and $J_\mathrm{G}$. As it can be seen, the current in the gate is significantly smaller than the current to the drain, thus making this motif an efficient photonic thermal switch. The thermal conductance is shown in Fig.~\ref{fig:Figure6}b, where we obtain the same qualitative response as with the asymmetric free-standing hBN thermal switch described in Figs.~\ref{fig:Figure4}b, d. However, unlike the case of free-standing hBN monolayers where the symmetric and anti-symmetric SPhP modes played similarly important roles in the transfer of radiative heat, when an hBN monolayer is placed atop of a substrate, like W, the symmetric SPhP mode vanishes due to the negligible thickness of monolayer hBN. Hence, in the spectral heat flux shown in Fig.~\ref{fig:Figure6}c, the low-frequency (near $0.16-0.17$ eV) resonances are drastically reduced with respect to Fig. \ref{fig:Figure5}. This happens because, in this case, these resonances represent the symmetric SPhP mode of the gate-hBN alone, while the symmetric SPhP modes of the drain and source have vanished. By contrast, the anti-symmetric SPhP resonances are preserved, hence we observe three resonances in the high-frequency end of the heat spectrum, corresponding to source, gate, and drain. Similar to the case of a free-standing asymmetric thermal switch, at $X_\mathrm{G}=0\%$ the SPhP mode of the gate is aligned with that of the source, whereas at $X_\mathrm{G}=1\%$ it is aligned with that of the drain, thereby creating the same symmetric dependence of the thermal conductance with respect to $X_\mathrm{G}=0.5\%$ as shown in Fig. \ref{fig:Figure6}b. Therefore, we see here that the basic principle of operation of the thermal switch is not significantly perturbed by the presence of a substrate.}

\par{To conclude, we proposed and analysed a theory for a three resonator-based thermal switch that operates based on the alignment and misalignment of ultra-narrow band photonic resonances. We implement this concept with hBN monolayers due to their very sharp and highly tunable LO phonon resonances in the mid-IR. We show that a triplet of hBN monolayers can achieve nearly $100\%$ tunability in thermal conductance in a deep subwavelength device. This concept can also be used to efficiently modulate a thermal current between bulk materials, as we showed in the case of W, where the tunability of thermal conductance reaches $70\%$. Our results pave the way for ultra-compact thermal switches based on radiative heat transfer in the near-field.}\\

\noindent \textbf{ORCID}\\
Georgia T. Papadakis: 0000-0001-8107-9221\\
Shanhui Fan: 0000-0002-0081-9732\\
 
\noindent \textbf{Notes}\\
The computational package used for near-field heat transfer calculations can be found in~\cite{CHEN2018163}. The authors declare no competing financial interest.  We acknowledge the support from the Department of Energy ``Photonics at Thermodynamic Limits'' Energy Frontier Research Center under Grant No. DE-SC0019140. G.T.P. acknowledges the TomKat Postdoctoral Fellowship in Sustainable Energy at Stanford University.\\

\section{\label{SUP}Supplemental Information}

{\subsection{Details on coupled-mode theory}

\par{In this section we present the details of the CMT formalism discussed in Section~\ref{Concept}. For the sake of generality, we replace the indices $S, G,$ and $D$ that pertained to the source, gate, and drain in Section \ref{Concept} with the indices $1, 2, $ and $3$, respectively. By writing the system of equations in Eq.~\ref{eq:1} in a matrix form, with $\boldsymbol{\alpha}=\{\alpha_\mathrm{1},\alpha_\mathrm{2},\alpha_\mathrm{3} \}^\intercal$, we have:
\begin{equation}\label{eq:SUP1}
d\boldsymbol{\alpha}/dt=i\Omega\boldsymbol{\alpha}+\Gamma\boldsymbol{\alpha}+\sqrt{2\Gamma}\boldsymbol{n}
\end{equation}
where $\boldsymbol{n}=[n_\mathrm{1},n_\mathrm{2},n_\mathrm{3}]^\intercal$, $\Gamma=\mathrm{diag} \{-\gamma_\mathrm{1},-\gamma_\mathrm{2},-\gamma_\mathrm{3}\}$, and $\Omega$ is given by
\begin{equation}\label{eq:SUP12}
\Omega=\left(\begin{array}{ccc} \omega_\mathrm{1} & \kappa_\mathrm{12} & \kappa_\mathrm{13} \\ \kappa_\mathrm{12}^{*} & \omega_\mathrm{2} & \kappa_\mathrm{23}  \\ \kappa_\mathrm{13}^{*}  & \kappa_\mathrm{23}^{*}  & \omega_\mathrm{3} \end{array}\right).
\end{equation}
The system of equations in Eq. \ref{eq:SUP1} can be solved to give: 
\begin{equation}\label{eq:SUP2}
\boldsymbol{\alpha}=\frac{1}{|F|} \left(\begin{array}{ccc} D_\mathrm{11} & D_\mathrm{12} & D_\mathrm{13} \\ D_\mathrm{21} & D_\mathrm{22} & D_\mathrm{23} \\ D_\mathrm{31} & D_\mathrm{32} & D_\mathrm{33} \end{array}\right) \sqrt{2\Gamma}\boldsymbol{n}.
\end{equation}
Hence, the mode amplitudes are given by:
\begin{equation}\label{eq:SUP3}
      \begin{aligned}
\alpha_\mathrm{1}=\frac{\sqrt{2}}{|F|}(\sqrt{\gamma_\mathrm{1}}D_\mathrm{11}n_\mathrm{1}+\sqrt{\gamma_\mathrm{2}}D_\mathrm{12}n_\mathrm{2}+\sqrt{\gamma_\mathrm{3}}D_\mathrm{13}n_\mathrm{3})\\
\alpha_\mathrm{2}=\frac{\sqrt{2}}{|F|}(\sqrt{\gamma_\mathrm{1}}D_\mathrm{21}n_\mathrm{1}+\sqrt{\gamma_\mathrm{2}}D_\mathrm{22}n_\mathrm{2}+\sqrt{\gamma_\mathrm{3}}D_\mathrm{23}n_\mathrm{3})\\
\alpha_\mathrm{3}=\frac{\sqrt{2}}{|F|}(\sqrt{\gamma_\mathrm{1}}D_\mathrm{31}n_\mathrm{1}+\sqrt{\gamma_\mathrm{2}}D_\mathrm{32}n_\mathrm{2}+\sqrt{\gamma_\mathrm{3}}D_\mathrm{33}n_\mathrm{3})\\
  \end{aligned}
  \end{equation}  
where:
\begin{equation}\label{eq:SUP4}
      \begin{aligned}
\left|F\right|=[i(\omega-\omega_\mathrm{1})+\gamma_\mathrm{1}][i(\omega-\omega_\mathrm{2})+\gamma_\mathrm{2}][i(\omega-\omega_\mathrm{3})+\gamma_\mathrm{3}]+\\
\left|\kappa_\mathrm{13}\right|^2[i(\omega-\omega_\mathrm{2})+\gamma_\mathrm{2}]+\left|\kappa_\mathrm{23}\right|^2[i(\omega-\omega_\mathrm{1})+\gamma_\mathrm{1}]+\\
\left|\kappa_\mathrm{12}\right|^2[i(\omega-\omega_\mathrm{3})+\gamma_\mathrm{3}]+i(\kappa_\mathrm{12}\kappa_\mathrm{23}\kappa_\mathrm{13}^{*}+\kappa_\mathrm{12}^{*}\kappa_\mathrm{23}^{*}\kappa_\mathrm{13}).
  \end{aligned}
  \end{equation}}
  \par{The elements of the tensor $D$ are given by:
\begin{equation}\label{eq:SUP5}
      \begin{aligned}
D_\mathrm{11}=(i\omega+\gamma_\mathrm{2}-i\omega_\mathrm{2})(i\omega+\gamma_\mathrm{3}-i\omega_\mathrm{3})+|\kappa_\mathrm{23}|^2 \\
D_\mathrm{21}=i\kappa_\mathrm{12}^{*}(i\omega+\gamma_\mathrm{3}-i\omega_\mathrm{3})-\kappa_\mathrm{23}\kappa_\mathrm{13}^{*} \\
D_\mathrm{31}=i\kappa_\mathrm{13}^{*}(i\omega+\gamma_\mathrm{2}-i\omega_\mathrm{2})-\kappa_\mathrm{12}^{*}\kappa_\mathrm{23}^{*} \\
D_\mathrm{12}=i\kappa_\mathrm{12}(i\omega+\gamma_\mathrm{3}-i\omega_\mathrm{3})-\kappa_\mathrm{13}\kappa_\mathrm{23}^{*} \\
D_\mathrm{22}=(i\omega+\gamma_\mathrm{1}-i\omega_\mathrm{1})(i\omega+\gamma_\mathrm{3}-i\omega_\mathrm{3})+|\kappa_\mathrm{13}|^2 \\
D_\mathrm{32}=i\kappa_\mathrm{23}^{*}(i\omega+\gamma_\mathrm{1}-i\omega_\mathrm{1})-\kappa_\mathrm{12}\kappa_\mathrm{13}^{*} \\
D_\mathrm{13}=i\kappa_\mathrm{13}^(i\omega+\gamma_\mathrm{2}-i\omega_\mathrm{2})-\kappa_\mathrm{12}\kappa_\mathrm{23} \\
D_\mathrm{23}=i\kappa_\mathrm{23}(i\omega+\gamma_\mathrm{1}-i\omega_\mathrm{1})-\kappa_\mathrm{13}\kappa_\mathrm{12}^{*} \\
D_\mathrm{33}=(i\omega+\gamma_\mathrm{1}-i\omega_\mathrm{1})(i\omega+\gamma_\mathrm{2}-i\omega_\mathrm{2})+|\kappa_\mathrm{12}|^2, \\
  \end{aligned}
  \end{equation}  
and the spectral flux to resonator $3$ (i.e. the drain) is given by~\cite{Hauss}:
  \begin{equation}\label{eq:SUP6}
J_\mathrm{3}(\omega)=\frac{1}{2\pi^2}\mathrm{Im}[\kappa_\mathrm{13}\langle{\alpha_\mathrm{1}^{*}\alpha_\mathrm{3}}\rangle+\kappa_\mathrm{23}\langle{\alpha_\mathrm{2}^{*}\alpha_\mathrm{3}}\rangle]. 
\end{equation}}

\par{Therefore, by making use of the mode orthogonality (Eq.~\ref{eq:2}), from Eq.~\ref{eq:SUP3} we obtain:
\begin{equation}\label{eq:SUP7}
      \begin{aligned}
\langle{\alpha_\mathrm{1}^{*}\alpha_\mathrm{3}}\rangle=\frac{4\pi}{\|F\|^2}(\gamma_\mathrm{1}D_\mathrm{11}^{*}D_\mathrm{31}\Theta(T_\mathrm{1})+\gamma_\mathrm{2}D_\mathrm{12}^{*}D_\mathrm{32}\Theta(T_\mathrm{2})+ \\ \gamma_\mathrm{3}D_\mathrm{13}^{*}D_\mathrm{33}\Theta(T_\mathrm{3})) \\
\langle{\alpha_\mathrm{2}^{*}\alpha_\mathrm{3}}\rangle=\frac{4\pi}{\|F\|^2}(\gamma_\mathrm{1}D_\mathrm{21}^{*}D_\mathrm{31}\Theta(T_\mathrm{1})+\gamma_\mathrm{2}D_\mathrm{22}^{*}D_\mathrm{32}\Theta(T_\mathrm{2}) \\ +\gamma_\mathrm{3}D_\mathrm{23}^{*}D_\mathrm{33}\Theta(T_\mathrm{3})).\\ 
  \end{aligned}
  \end{equation}
From Eq.~\ref{eq:SUP7} one can see that the thermal exchange between resonators $1$ and $3$ is mediated by the temperature of resonator $2$ and, similarly, the thermal exchange between resonators $2$ and $3$ is mediated by the temperature of resonator $1$.}
    
\par{Next, to obtain the results of Eq. \ref{eq:6}, \ref{eq:7}, \ref{eq:8}, we assume that $\kappa_\mathrm{i,j}\ll\omega_\mathrm{i}$, for $i=1,2,3$. We further assume a symmetric switch, for which $\omega_\mathrm{1}=\omega_\mathrm{3}$ and $\gamma_\mathrm{1}=\gamma_\mathrm{3}$, $\kappa_\mathrm{12}=\kappa_\mathrm{23}$. In this case, Eqs. \ref{eq:SUP4}-\ref{eq:SUP7} yield:
\begin{equation}\label{eq:SUP8}
  \begin{aligned}
J_\mathrm{3}(\omega)=\frac{2\gamma_\mathrm{1}}{\pi}\{\frac{\kappa_\mathrm{13}^2\gamma_\mathrm{1}\Delta\Theta(\omega,T_\mathrm{1},T_\mathrm{3})}{(\delta\omega_\mathrm{1}^2+\gamma_\mathrm{1}^2)^2}
+ \frac{\kappa_\mathrm{23}^2\gamma_\mathrm{2}\Delta\Theta(\omega,T_\mathrm{2},T_\mathrm{3})}{(\delta\omega_\mathrm{1}^2+\gamma_\mathrm{1}^2)(\delta\omega_\mathrm{2}^2+\gamma_\mathrm{2}^2)} \\
+\frac{\kappa_\mathrm{23}^2\kappa_\mathrm{13}(\gamma_\mathrm{1}\delta\omega_\mathrm{2}-\gamma_\mathrm{2}\delta\omega_\mathrm{1})\Theta(\omega,T_\mathrm{1})}{(\delta\omega_\mathrm{1}^2+\gamma_\mathrm{1}^2)^2(\delta\omega_\mathrm{2}^2+\gamma_\mathrm{2}^2)})\} 
 \end{aligned}
\end{equation}
where $\Delta\Theta(\omega,T_\mathrm{i},T_\mathrm{j})=\Theta(\omega,T_\mathrm{i})-\Theta(\omega,T_\mathrm{j})$, and $\delta\omega_\mathrm{i}=\omega-\omega_\mathrm{i}$, for $i=1,2$. We note here that Eq. \ref{eq:SUP8} reduces to the result discussed in detail in \cite{Fan_RectivicationVacuumPRL2010} if we consider the interaction between two resonators instead of three, namely if we set $\kappa_\mathrm{13}=0$. The results of Eqs. \ref{eq:6b}-\ref{eq:9} pertain to the integration of $J_\mathrm{3}(\omega)$ of Eq. \ref{eq:SUP8}, for $\gamma_\mathrm{1}=\gamma_\mathrm{2}$.}

\par{In Fig. \ref{fig:Figure3}b we showed the agreement between the CMT formalism and the fluctuational electrodynamics formalism. The coupling rates as well as the resonant frequencies and loss rates were determined via curve-fitting of the CMT model to the result with fluctuational electrodynamics. These are: $\omega_\mathrm{S,s}=0.17$ eV, $\gamma_\mathrm{S,s}=0.151$ meV, $\omega_\mathrm{G,s}=0.166$ eV, $\gamma_\mathrm{G,s}=0.005$ meV, $\omega_\mathrm{D,s}=0.163$ eV, $\gamma_\mathrm{D,s}=0.107$ meV, $\kappa_\mathrm{SG,s}=\sqrt{\gamma_\mathrm{S,s}\gamma_\mathrm{G,s}}/16.7$, $\kappa_\mathrm{GD,s}=\sqrt{\gamma_\mathrm{G,s}\gamma_\mathrm{D,s}}/7.4$, $\kappa_\mathrm{SD,s}=\sqrt{\gamma_\mathrm{D,s}\gamma_\mathrm{S,s}}/1.12$, $\omega_\mathrm{S,a}=0.192$ eV, $\gamma_\mathrm{S,a}=0.076$ meV, $\omega_\mathrm{G,a}=0.189$ eV, $\gamma_\mathrm{G,a}=0.003$ meV, $\omega_\mathrm{D,a}=0.186$ eV, $\gamma_\mathrm{D,a}=0.164$ meV, $\kappa_\mathrm{SG,a}=18.35\sqrt{\gamma_\mathrm{S,a}\gamma_\mathrm{G,a}}$, $\kappa_\mathrm{GD,a}=4.84\sqrt{\gamma_\mathrm{G,s}\gamma_\mathrm{D,a}}$, $\kappa_\mathrm{SD,a}=11.5\sqrt{\gamma_\mathrm{S,a}\gamma_\mathrm{D,a}}$, where the subscripts ``s'' and ``a'' correspond to the symmetric and anti-symmetric modes, respectively.}
 
\subsection{Ab initio calculations}

\par{First principles calculations were performed to compute the dielectric function of monolayer hBN as a function of strain. Specifically, these calculations were based on density functional theory (DFT) and used a combination of Quantum ESPRESSO~\cite{qe1,qe2} and JDFTx~\cite{JDFTx}. All calculations used the PBEsol~\cite{PBEsol} exchange-correlation functional and corresponding ultrasoft pseudopotentials~\cite{USPP}. The unstrained in-plane lattice constant was 2.505 \AA. The phonon frequencies and eigenmodes were computed within Quantum ESPRESSO using density functional perturbation theory (DFPT)~\cite{sohier_dfpt_2017}, particularly with appropriate handling of the 2D Coulomb interaction present in these monolayer systems~\cite{sohier_LOTO_2017}. DFT calculations were performed using a plane-wave basis set with a kinetic energy cutoff of 40 Hartrees on a $24\times24\times1$ k-point grid. The phonons were modeled using a $6\times6\times1$ q-point mesh. Meanwhile, the phonon linewidths were computed using JDFTx along with Phono3py~\cite{phono3py1,phono3py2}, which uses finite differences to compute the anharmonic force constants due to three-phonon interactions, and also computes the effects of isotope scattering~\cite{isotope}. DFT calculations were performed here using JDFTx with a plane-wave basis set with a 40 Hartree kinetic energy cutoff. Electronic properties were computed using a $24\times24\times1$ k-point mesh and throughout Coulomb truncation was included~\cite{CoulombEXX} to ensure no artificial LO-TO splitting at the zone center. To capture the phonon properties, a $6\times6\times1$ supercell is used. To converge the phonon linewidths, a $150\times150\times1$ grid of q-points were used in Phono3py in conjunction with the tetrahedron integration method. }

\bibliographystyle{ieeetr}


\end{document}